\documentclass{llncs}
\input{psfig.sty}

\usepackage{graphicx}
\usepackage{psfrag}
\usepackage{amsmath}

\begin{document}

%\pagestyle{headings}
%In order to omit page numbers and running heads
%please change this line to
\pagestyle{empty}

%and change the first command line too, see above.

\mainmatter

\title{Topology of Cell-Aggregated Planar Graphs}

\titlerunning{Cell-Aggregation, Paths \& Cycles}

\author{ Milovan \v Suvakov and Bosiljka Tadi\'c }

\authorrunning{Milovan \v Suvakov \& Bosiljka Tadi\'c}

\institute{Department for Theoretical Physics, Jo\v zef Stefan Institute, Box 3000,
1001 Ljubljana, Slovenia\\\email{Milovan.Suvakov@ijs.si};
\email{Bosiljka.Tadic@ijs.si} \\ 
\texttt{http://www-f1.ijs.si/$\sim$tadic/}}

\maketitle

%Published in: Lecture Notes in Computer Science, V.N. Alexandrov
%{\it et al.} Eds.,\\ Springer (Berlin) Part III, 3993 (2006) 1098-1105

\begin{abstract}
We present new algorithm for growth of non-clustered planar graphs by aggregation of cells with given distribution of size and constraint of  connectivity $k = 3$ per node. The emergent graph structures are controlled by two parameters---chemical potential of the cell aggregation and the width of the cell size  distribution. We compute several statistical properties of these 
graphs---fractal dimension of the perimeter, distribution of shortest paths between pairs of nodes and topological betweenness of nodes and links. We show how these topological properties depend on the control parameters of the aggregation process and discuss their relevance for  the 
conduction of current in self-assembled nanopatterns.
\end{abstract}

\section{Introduction}
In recent years increased interests in various networks realizations 
\cite{SD_book,nets_review}  revealed
that  several new types of graphs termed {\it structured graphs} are more appropriate mathematical objects to describe complex network's geometry than traditional {\it random graphs} \cite{BB_book}. The variety of structures was found to emerge through evolution processes in which nodes and links are added sequentially according to specified rules, in particular, the preferential attachment rules lead to strongly inhomogeneous {\it scale-free graphs} \cite{SD_book}. In contrast to the evolving networks, which comprise a class of {\it causal} graphs, the class of {\it homogeneous} graphs consists of graphs with fixed number of nodes and fluctuating or rewiring links according to given rules or 
certain optimization processes. Complex graph structures may emerge in this procedures, especially when certain global or local optimization constraints are imposed \cite{Stefan}.

Planar graphs are special class of graphs that can be embedded in a Euclidean plane. A graph is planar {\it iff it does not contain a subdivision of $K_5$ (5-clique) and $K_{3,3}$ (minimal non-planar graph with 6 nodes)} \cite{BB_book}. Consequently, planar graphs fulfill Euler's law: $N_p+N=E+1$, which is relating the number of nodes $N$, links $E$ and 
polygons $N_p$.

In this work we suggest a new method for growing a planar cellular graph
by attachment of objects---cells (polygons) of length $n_p$, which are chosen from a given distribution $f(n_p)$. The polygons are added sequentially in time starting from an initial polygon. In addition, we strictly impose the constraint on number of links per node $k=3$, which is thus fulfilled everywhere in the interior of the graph and on some nodes on the graph boundary.
The attachment of cells is controlled by two parameters---the width of the distribution of cell sizes $\mu_2$ and the parameter $\nu$  that plays the role of chemical potential of cell aggregation. In the limit of vanishing attachment potential $\nu \to 0$ the growth process resembles the one in  diffusion-limited aggregation \cite{DLA}. However, aggregated are spatially extended cells of particles rather than single particles.

Emergent structures of cellular networks are resembling of soap froths \cite{Stabans_froth} or patterns of nano-particles self-assembled through  nonlinear dynamic processes \cite{SA_book,Philip_SA}. Typically, a pattern of cells appears  when nano-particles are immersed in a liquid film, which is then allowed to evaporate
until holes of different sizes open-up leaving particles in the walls between
the holes \cite{Philip_SA,Philip_REP}.  Generally, the structure of the patterns effects the physical processes on them, such as current transport \cite{MSetal}. It is therefore important to understand the topology of the aggregated cellular networks in detail.
Here we study the topological properties, such as shortest paths between nodes, topological centrality,  and fractality of the graph's perimeter in different cellular networks obtained by varying the control parameters of the aggregation processes.

\section{Cell Aggregation}
The basic idea is to make growing model of planar graph with given distribution of cell (polygon) sizes $f(n_p)$ and with degree of nodes $2 \le k \le 3$.

\subsection{Topological constraint}
For this purpose we observe some topological constraints on the
distribution of cell sizes: (i) $f(n_p)$ is defined for $n_p \geq 3$, 
for non-clustered graph we fix $f(3)=0$; (ii) planar graph obeys Euler's law: $N_p+N=E+1$.
Among these the homogeneous plane-filling structures are of special interest \cite{Stabans_froth}. For this class of graphs majority of nodes are in the interior of the graph, ie., nodes with degree $k=3$. 
Therefor $3N \approx 2E$ and Euler's law becomes
\begin{equation} \label{Euler1}
6N_p=2E+6.
\end{equation}
For large system with distribution of cell sizes $f(n_p)$ we have
\begin{equation}
N=N_p\sum_{n_p} \frac{n_pf(n_p)}{3},\; E=N_p\sum_{n_p} \frac{n_pf(n_p)}{2}. 
\end{equation}
Substituting second relation into (\ref{Euler1}): $6N_p=N_p\sum_{n_p} n_pf(n_p)+6$,
then for large $N_p \gg 1$ one can find  that the average cell size is 
equal to six
\begin{equation}\label{tc}
\langle n_p \rangle \equiv \sum_{n_p} n_pf(n_p)=6.
\end{equation}

We use lognormal distribution of polygon size 
\begin{equation}\label{lnormal}
f(n_p) = \frac{1}{s \sqrt{2\pi} x} e^{-\frac{\ln^2{x/x_0}}{2s^2}},
\end{equation}
which is most often found in experiments \cite{Philip_SA}. Using the condition (\ref{tc}) the number of independent parameters in (\ref{lnormal})
is reduced
\begin{equation}
\langle n_p \rangle = 6 \quad \Rightarrow \quad x_0=6e^{-s^2/2}, \; s^2 = \ln \left( 1+ \frac{\mu_2}{36} \right),
\end{equation}
where second central moment  $\mu_2$ remains as the control parameter in our case.

\subsection{Model}
Starting from an initial cell,
at each time step a new cell with size taken from the distribution $f(n_p)$ is attached to the graph boundary of the graph according to the following rules:
\begin{enumerate}
\item In the aggregation only nodes with degree 3 can be closed inside the graph;
\item New link can be assigned only to nodes with degree 2;
\item Potential nesting place is searched as an array of nodes on graph boundary with degree 3 limited with two nodes of  degree 2. The nested part of the cell is identified with the nodes of the nesting string.
Therefore, number of extra nodes to be added is $n=n_p-l$, where $l$ is length of the nesting string. We select the nesting place with probability $p \sim e^{-\nu n}$, where the parameter $\nu$ plays the role of the chemical potential for addition of new nodes.
\end{enumerate}

\begin{figure}[!h]
\begin{center}
$\begin{array}{c@{\hspace{0.1in}}c}
\includegraphics[width=0.48\textwidth]{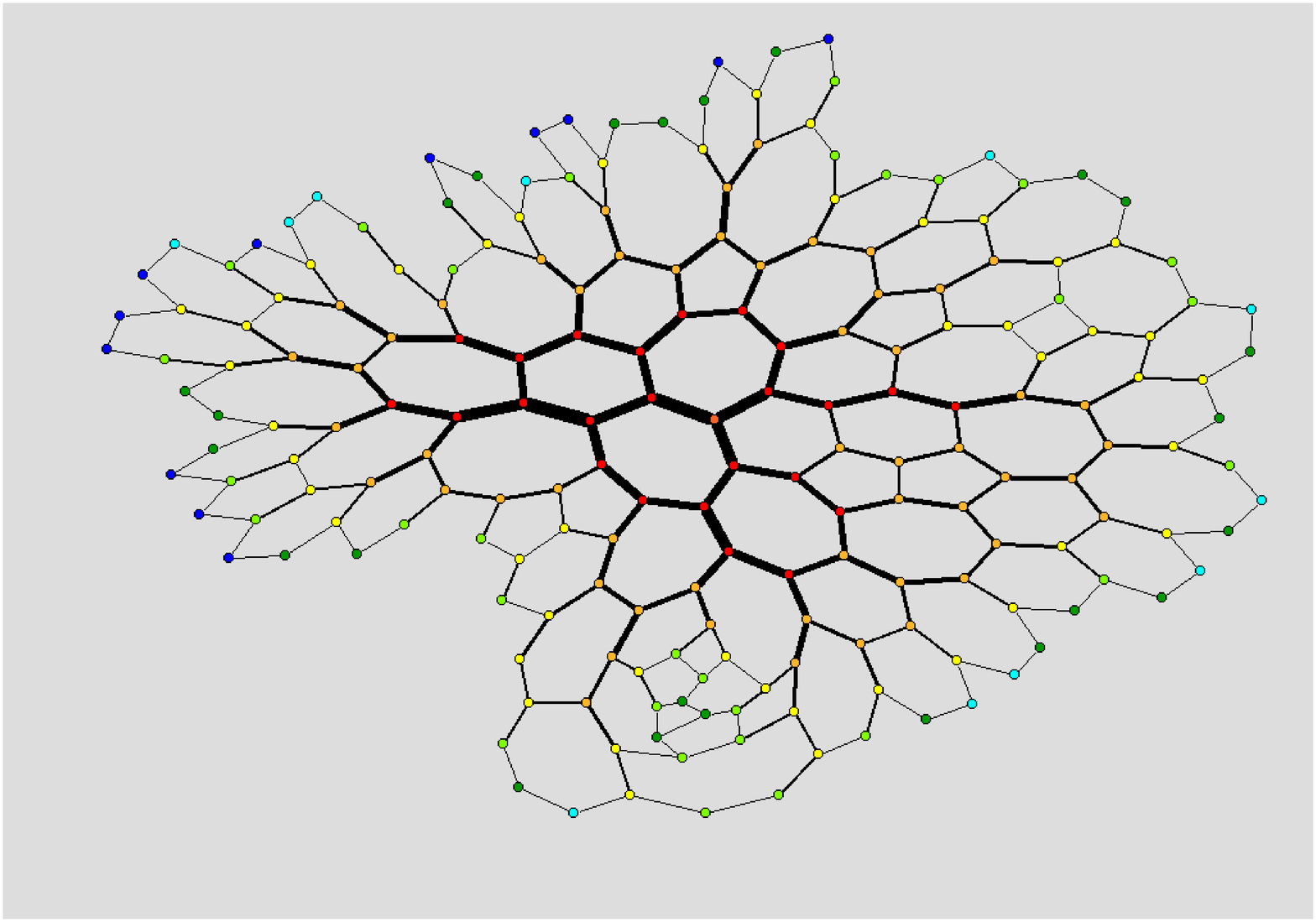} &
\includegraphics[width=0.48\textwidth]{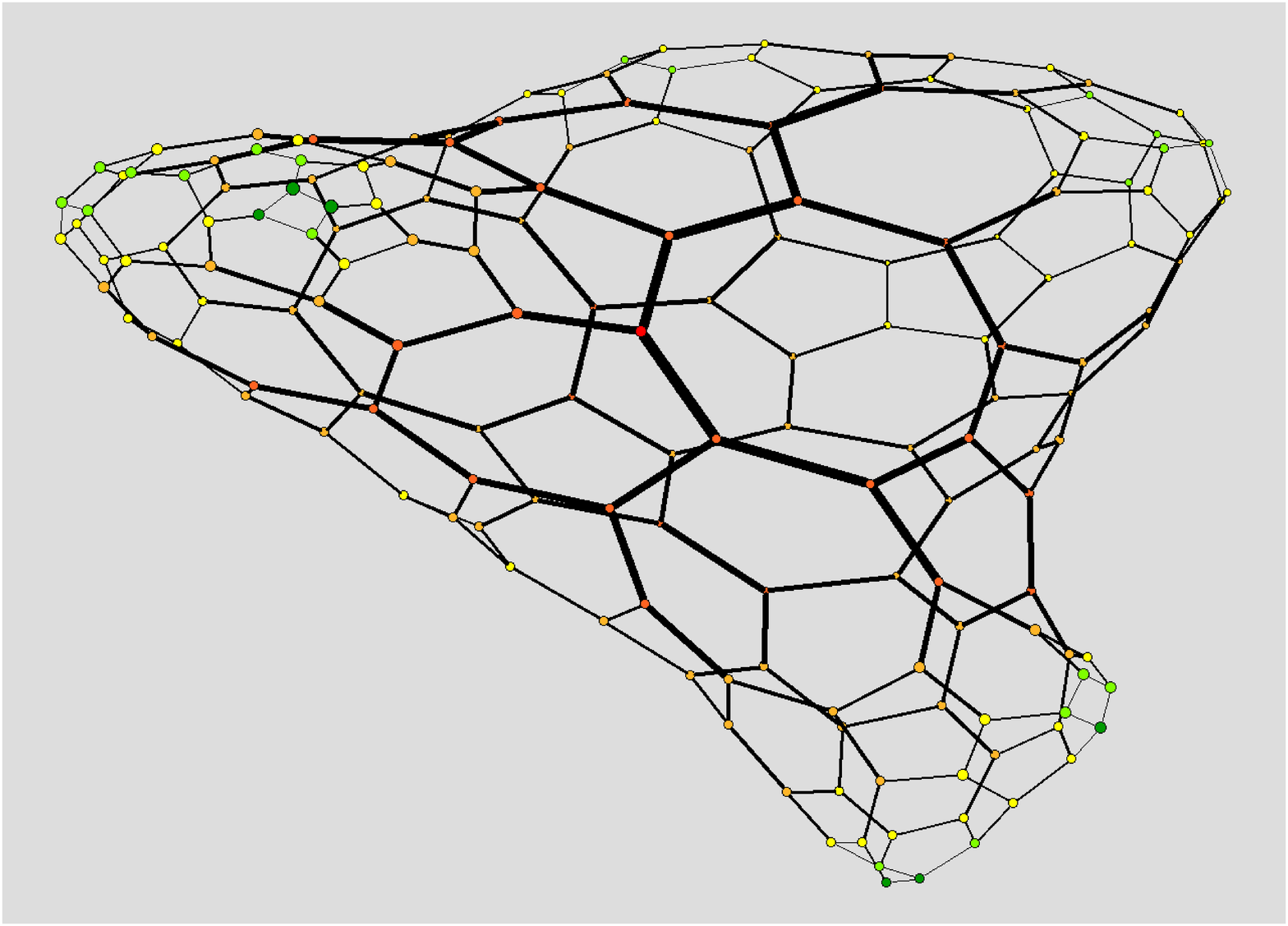}\\
\mbox{Open} & \mbox{Closed}
\end{array}$
\end{center}
\caption{Two possible types of cell-aggregated \textit{planar graphs}: open and closed structures, obtained by lognormal distribution with $\mu_2=1.0$, and aggregation potential  $\nu=5.0$. }
\label{ozslika}
\end{figure}

\newpage

We implemented this algorithm in \textit{C++} according to following steps:
\begin{verbatim}
Initial graph: one polygon of size np taken from f(np) 
For i=2 to Np
  np = next random from distribution f(np)
  If(there is no nodes on graph boundary with degree 2) exit(1)
  For all j=(periphery node with degree 2)
    d=distance to the next node on graph boundary with degree 2
    Number of new nodes n = np-d-1
    If(n>0) p(j)=exp(-nu*n)
  End of loop j
  If(there is no j with n>0) exit(2)
  Normalize p(j)
  j = next random from distribution p(j)
  Add new polygon with size np linked with 
    node j and next node on graph boundary with degree 2
End of loop i
exit(0)
\end{verbatim}

\begin{figure}[!h]
\begin{center}
$\begin{array}{c@{\hspace{0.1in}}c@{\hspace{0.1in}}c}
\mbox{$\mu_2$} & \mbox{$\nu=0.0$} & \mbox{$\nu=1.0$} \\
\mbox{$0.5$} &
\includegraphics[width=0.45\textwidth]{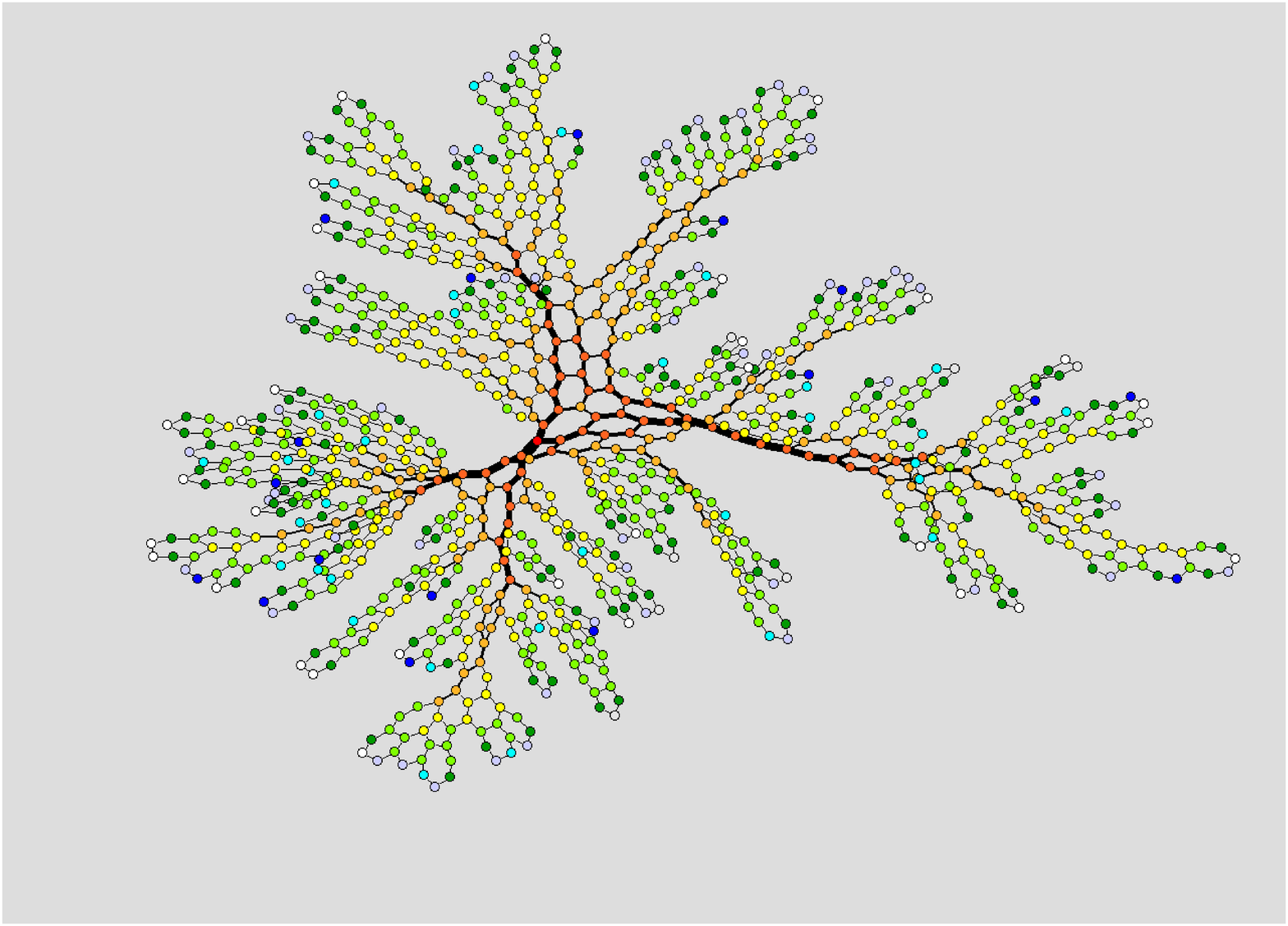} &
\includegraphics[width=0.45\textwidth]{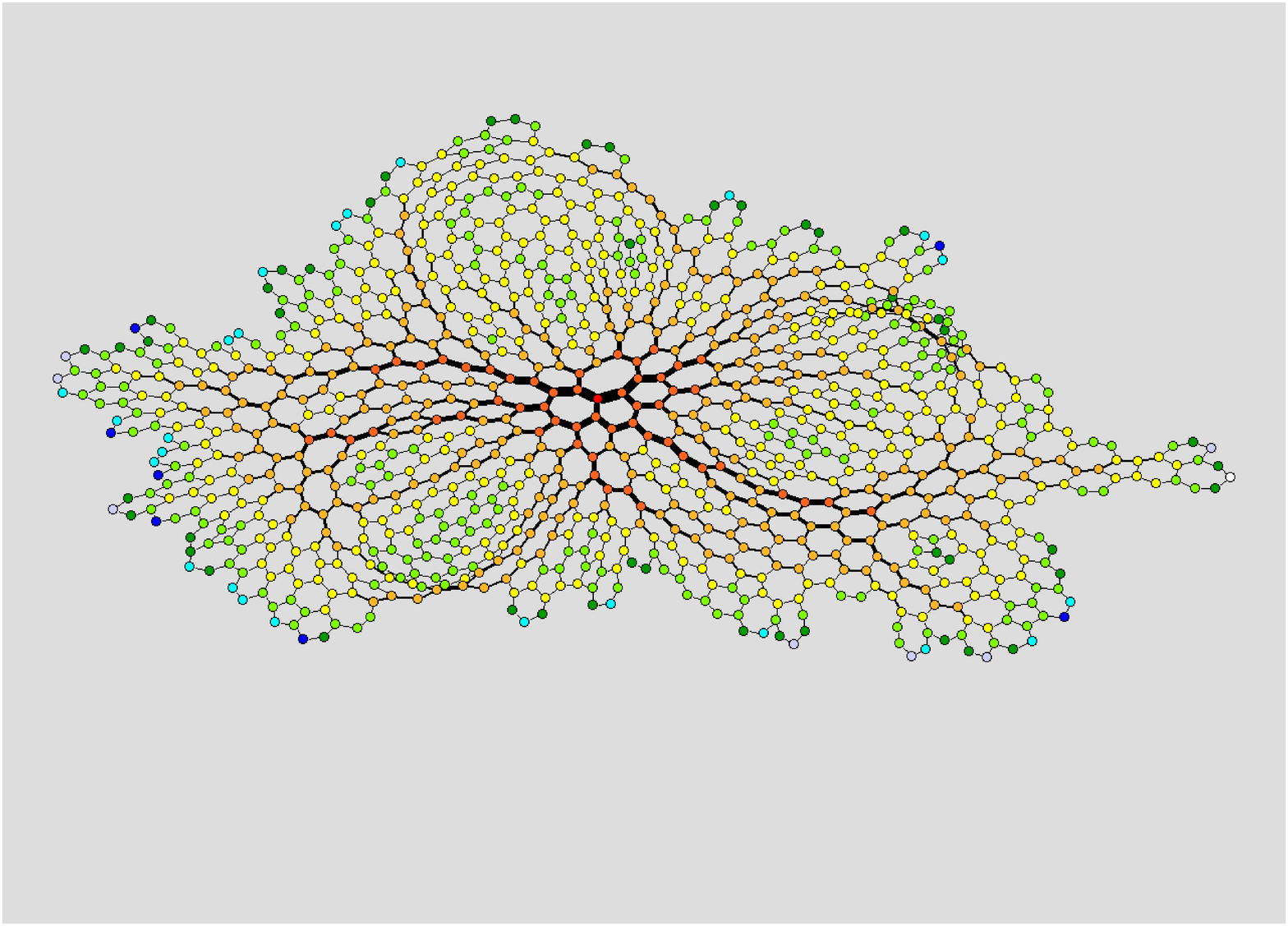} \\
\mbox{$2.0$} &
\includegraphics[width=0.45\textwidth]{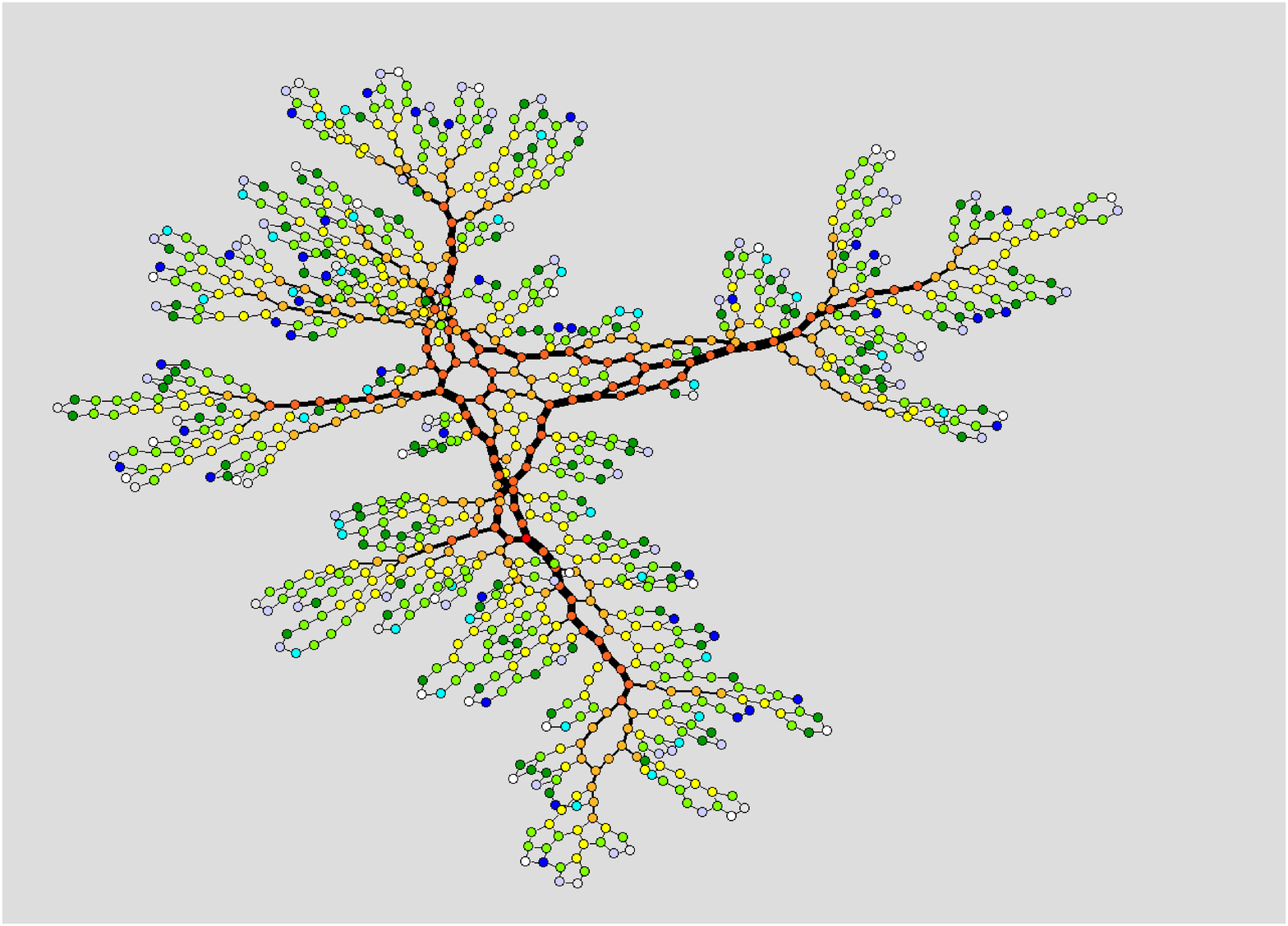} &
\includegraphics[width=0.45\textwidth]{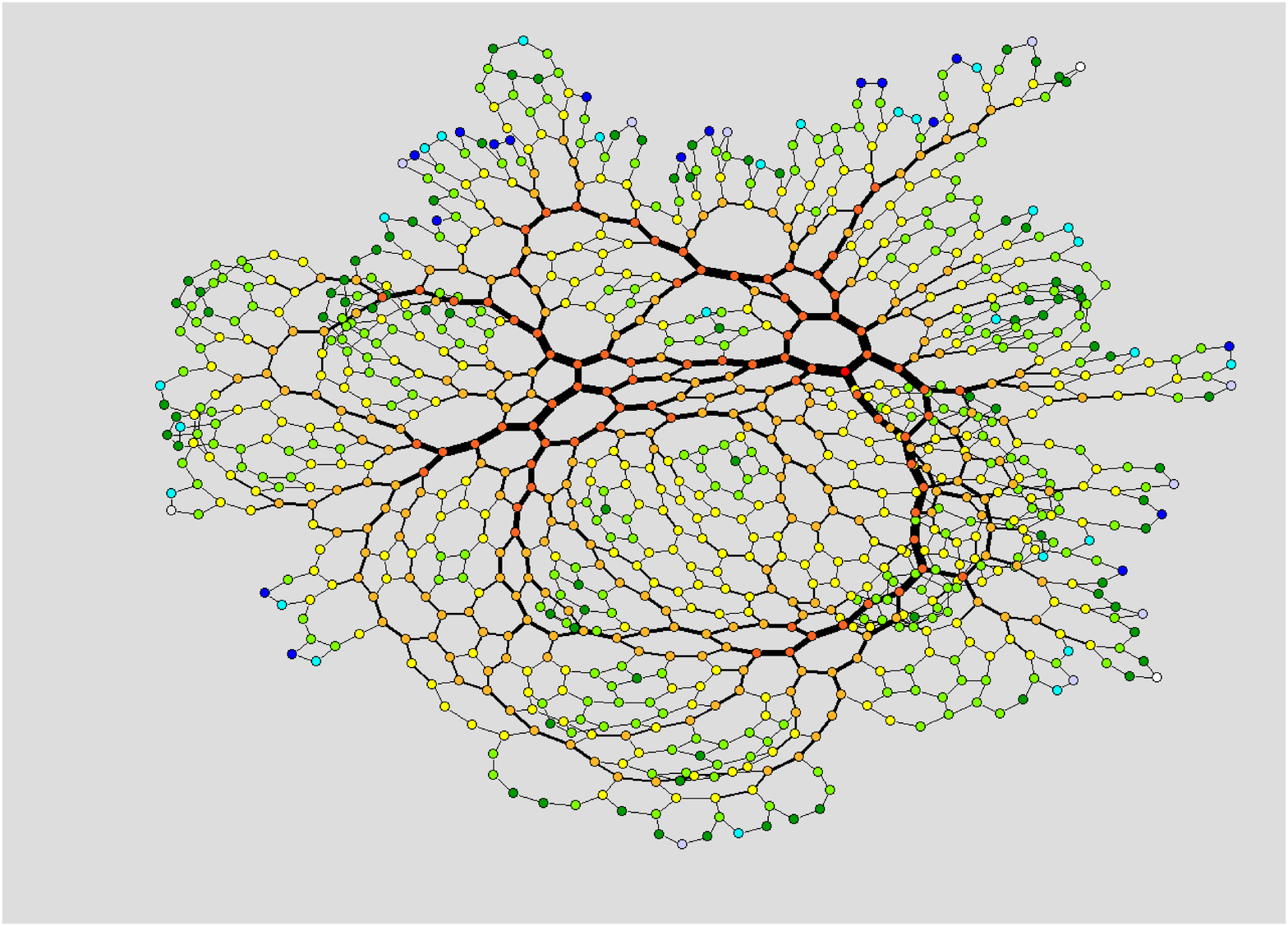} \\
\end{array}$
\end{center}
\caption{Cell-aggregated planar graphs with lognormal distribution of polygon size for various values of width $\mu_2$ and cell aggregation potential $\nu$. Width of lines represent topological betweenness (centrality) of links calculated  in Section \ref{centrality}. }
\label{velikaslika}
\end{figure}

Depending on model parameters of the growth process and it stochasticity three possible exit cases are:
\begin{itemize}
\item exit(0) - Open structure (planar graph with $N_p$ polygons);
\item exit(1) - Closed structure (after some number of step there are no more nodes of degree 2 and structure stops to grow,  no nesting places of any size);
\item exit(2) - No nesting place available for current cell. In this case one can take next cell, which in turn perturbs the
actual  distribution.
\end{itemize}
We never experienced the exit(2) situation for the range of parameters $\mu_2 \in [0.5, 2.0]$, $\nu \in [0,5]$ and $N_p=1000$ in huge number of samples. Two examples of the emergent open and closed structures are shown on Fig. \ref{ozslika}.
More examples of cellular networks are shown in Fig. \ref{velikaslika} for varying parameters $\mu_2$ and $\nu$..

\section{Fractal Dimension of Network Perimeter}
\begin{figure}[!h]
\begin{center}
%$\begin{array}{cc}
\includegraphics[width=1.0\textwidth]{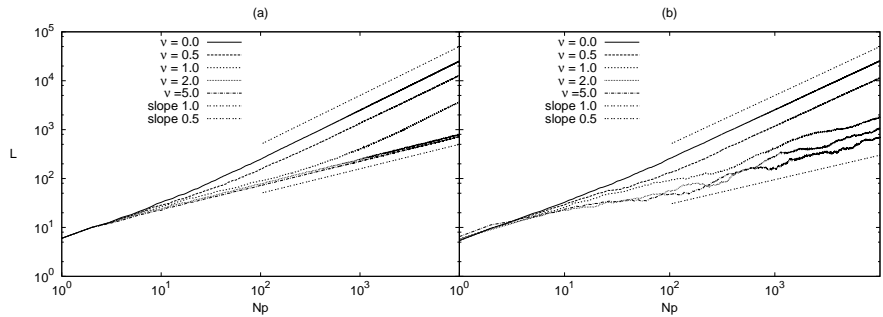} 
%&
%\includegraphics[width=0.5\textwidth]{dln2.0fd.eps} \\
%\mbox{(a)} & \mbox{(b)}
%\end{array}$
\end{center}
\caption{Scaling of the network perimeter for (a) $\mu_2=0$ hexagons only (b) $\mu_2=2.0$}
\label{FD}
\end{figure}

During the nesting growing process in one step number of nodes $N$ increases by $n<n_p$. Number of added nodes $n$ at each step depends on the cell size and  length of the nesting string. Therefore, $N \approx \kappa N_p$ where $\kappa=\langle n \rangle$ is average growth rate. For open structures (see Fig. \ref{ozslika}-\ref{velikaslika}) boundary of the graph becomes fractal, depending on the control parameters.
In fact, length of the graph boundary $L$ grows as a power of the number of cells $N_p$ (or network size $N$) with fractal dimension $D$ defined by $L \sim N_p^D$.

In Fig. \ref{FD} we show how number of nodes on the graph boundary increases with $N_p$. Each point is averaged over $10$ emergent growing networks. The dimension $D$ is in the range $\frac{1}{2} \le D \le 1$, when $D=1$ correspond to structures of high fractality, that is obtained for small values of the parameter $\nu$. $D=\frac{1}{2}$ correspond to planar "circle like" structures with reduced fractality. For $\mu_2>0$ we observe a continuous crossover between these two limits (see Fig. \ref{FD}b for $\mu_2=2$). However, in structures with homogeneous cell distribution ($\mu_2=0$ - hexagons only) a sharp transition seems to occur at $\nu_c \approx 1.5$.

\section{Shortest Paths and Centrality on Cellular Networks}

In this section we consider global topological properties of the cell-aggregated planar graphs and their dependence on the control parameters $\mu_2$ and $\nu$.

\subsection{Shortest Paths}

\begin{figure}[!h]
\begin{center}
{\psfig{file=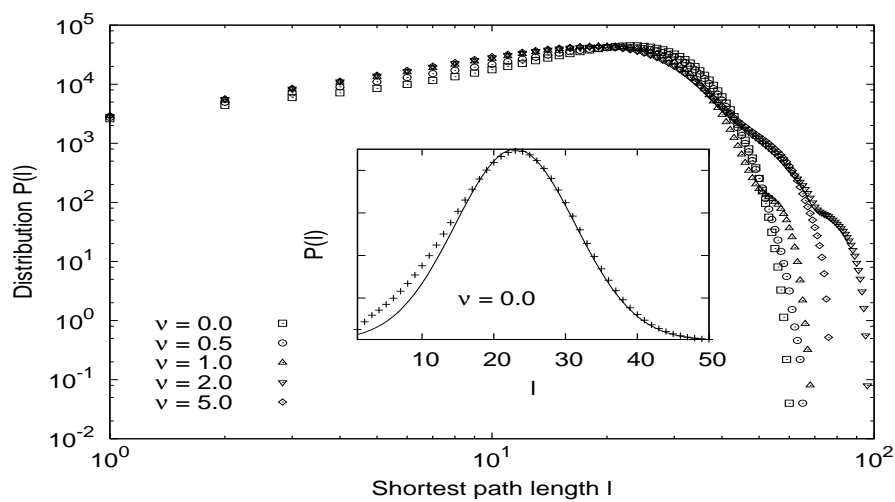,height=2.6in,width=4.8in}}
\end{center}
\caption{Distribution of lengths of shortest paths on networks for fixed $\mu_2=1.0$ and various values of $\nu$.
Insert: The distribution in the case $\nu=0$ is shown on  linear scale. Solid line: Gaussian with $l_0=23$ and $\sigma=8.37$.}
\label{DSP}
\end{figure}

Shortest path between two nodes is defined as path along the smallest number of intermediate links \cite{BB_book}. We implemented an algorithm for counting shortest paths of Dijksta type \cite{NW_book}. In Fig. \ref{DSP} we show distribution of lengths of shortest paths between all pairs of nodes on network. All networks are for fixed $\mu_2=1.0$ and approximately of the same size 
$N \approx 1000$ nodes. Each point in Fig. \ref{DSP} is averaged over 100 sample networks. We found similar results  for other $\mu_2$ values.

All these networks have similar topology at local level, because the
number of links at all interior nodes is constant $k=3$. Therefore, distributions of shortest distances at small scale are similar for all values of parameter $\nu$. 
Differences in global topology appear on large scale for lengths larger then peak value $l_0 \sim 25$, which manifest in occurrence of additional peaks (see Fig. \ref{DSP}). The  probability of long paths increases for larger values of the parameter $\nu$. Whereas  in the limit $\nu=0$ the distribution of 
length of shortest paths  on large scale can be approximated with a normal distribution (inset on Fig. \ref{DSP}).

\subsection{Centrality Measures} \label{centrality}

\begin{figure}[!h]
\begin{center}
{\psfig{file=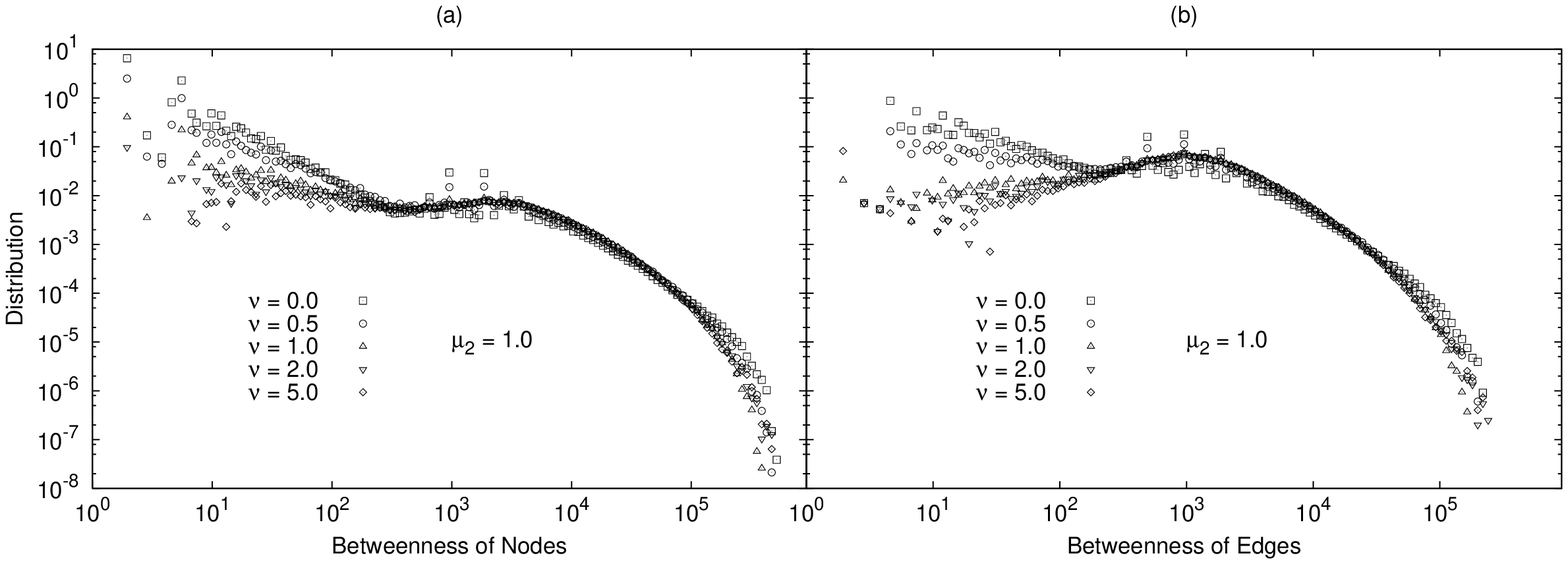,height=2.6in,width=4.8in}}
\end{center}
\caption{Distribution of betweenness (a) of nodes (b) of links.}
\label{DBC}
\end{figure}

Betweenness centrality of a node in network is defined by \cite{Freeman,BB_book}
\begin{equation}
C_B(v)=\sum_{s\neq v \neq t} \frac{\sigma_{st}(v)}{\sigma_{st}}
\end{equation}
where $\sigma_{st}$ is total number of shortest paths between nodes $s$ and $t$, and $\sigma_{st}(v)$ is number of these paths that node $v$ lies on. Betweenness of links is defined in analogous way.  In our algorithm, we record number of shortest paths through each node and through each link on a network. In Fig. \ref{DBC} we show distributions of betweenness of nodes and links averaged over 100 sample networks with fixed $\mu_2=1.0$ and size $N \approx 1000$ nodes.
For inhomogeneous networks, distributions of these two betweenness measures can be substantially different. In our case, however, they are similar because all interior nodes have fixed degree $k=3$.
We find that distributions at small scale strongly depend on parameter $\nu$, which results from the graph boundary.
Similar conclusions hold for other $\mu_2$ values.

In Fig. \ref{velikaslika} thick lines represent links with large betweenness. For this type of networks the strongest lines, which make the skeleton of the graph,   are connecting the nodes with largest centrality measure.

\section{Conclusions}

We have introduced a new algorithm for growth of graphs by aggregation process of extended objects - polygons with size distribution. Depending on aggregation conditions, which are determined by  two parameters controlling the
distribution width and attachment potential, we can get a wide spectrum of 
emergent structures. 
In this paper we presented some results for the case of lognormal distribution of cells and additional constraints, leading to the emergent non-clustered  planar graphs with a constant node connectivity.
 The algorithm works for variety of cell distributions and constraints, that may result in diverse opened (fractal) or closed structures. For instance, 
for a special set of parameters we can get closed structures of  $C_{60}$ type.

We measured several topological properties of these networks in quantitative details---fractality of the graph boundary, shortest paths, and betweenness centrality. This properties are important for some dynamical process on networks such as electrical conductivity \cite{MSetal} via single electron tunneling \cite{Philip_SA,Nano_book}.

{\bf Acknowledgments:}\\
M.S. thanks financial support from the Marie Curie Research and Training Network   MRTN-CT-2004-005728 project. B.T. is supported by the program
P1-0044 of the Ministry of high education, science and technology (Slovenia).

\end{document}